\documentclass[journal]{IEEEtran}
\usepackage[latin9]{inputenc}
\usepackage{geometry}
\usepackage{color}
\usepackage{float}
\usepackage{amsmath}
\usepackage{amssymb}
\usepackage{graphicx}

\makeatletter

\floatstyle{ruled}
\newfloat{algorithm}{tbp}{loa}
\providecommand{\algorithmname}{Algorithm}
\floatname{algorithm}{\protect\algorithmname}

\usepackage{amsfonts}
\usepackage{mathrsfs}
\usepackage{mathrsfs}\usepackage[noblocks]{authblk}
\usepackage[compress]{cite}
\usepackage{bm}
\usepackage{algorithm}
\usepackage{algorithmic}
\usepackage{epsfig}\usepackage{graphics}\usepackage{subfigure}\usepackage{epsfig}\usepackage{epstopdf}
\usepackage{graphics}\usepackage{subfigure}\usepackage{upgreek}\usepackage{caption}\usepackage{theorem}
\usepackage{breqn}
\usepackage{multicol}
\usepackage{eufrak}
\usepackage{eucal}
\usepackage{array}
\usepackage[compress]{cite}
\usepackage{algorithm}
\usepackage{algorithmic}

\newtheorem{lemma}{Lemma}\theoremheaderfont{\normalfont\bfseries}

\makeatother

\begin{document}
\title{Robust Beamforming Design for Intelligent Reflecting Surface Aided
MISO Communication Systems}
\author{Gui~Zhou, Cunhua~Pan, Hong~Ren, Kezhi~Wang, Marco~Di~Renzo,~\IEEEmembership{Fellow,~IEEE,}
and Arumugam~Nallanathan,~\IEEEmembership{Fellow, IEEE}
 \thanks{G. Zhou, C. Pan, H. Ren and A. Nallanathan are with the School of
Electronic Engineering and Computer Science at Queen Mary University
of London, London E1 4NS, U.K. (e-mail: g.zhou, c.pan, h.ren, a.nallanathang@qmul.ac.uk).
K. Wang is with Department of Computer and Information Sciences, Northumbria
University, UK. (e-mail: kezhi.wang@northumbria.ac.uk). M. Di Renzo
is with Université Paris-Saclay, CNRS, CentraleSupélec, Laboratoire
des Signaux et Systemes, Gif-sur-Yvette, France (e-mail: marco.direnzo@centralesupelec.fr).}}
\maketitle
\begin{abstract}
Perfect channel state information (CSI) is challenging to obtain due
to the limited signal processing capability at the intelligent reflection
surface (IRS). This is the first work to study the worst-case robust beamforming
design for an IRS-aided multiuser multiple-input single-output (MU-MISO)
system under the assumption of imperfect CSI. We aim for minimizing
the transmit power while ensuring that the achievable rate of each
user meets the quality of service (QoS) requirement for all possible
channel error realizations. With unit-modulus and rate constraints,
this problem is non-convex. The imperfect CSI further increases the
difficulty of solving this problem. By using approximation and transformation
techniques, we convert this problem into a squence of semidefinite
program (SDP) subproblems that can be efficiently solved. Numerical
results show that the proposed robust beamforming design can guarantee
the required QoS targets for all the users. 
\end{abstract}


\section{Introduction}

Intelligent reflecting surface (IRS) has recently been proposed as
a cost-effective and energy-efficient high data rate communication
technology due to the rapid development of radio frequency (RF) micro-electro-mechanical
systems (MEMS) as well as the abundant applications of the programmable
and reconfigurable metasurfaces \cite{EURASIP2019}. It consists of
an passive array structure that is capable of adjusting the phase
of each passive element on the surface continuously or discretely
with low power consumption \cite{Marco-3,Marco-4}. The benefits of
IRS in enhancing the spectral and energy efficiency have been demonstrated
in various schemes (e.g., \cite{Pan2019intelleget,Xianghao2009,Gui2019IRS,OFDM2019,Baitong2019,Marco-2})
by the joint design of active precoder at the base station (BS) and
passive reflection beamforming at the IRS.

However, all the existing contributions on IRS are based on the assumption
of perfect channel state information (CSI) at the BS, which is too
idealistic in IRS communications. For the imperfect
CSI, the authors in \cite{Jung2020TWC} studied the impact of the
channel error by adopting the performance analysis technique in an
uplink MISO system. There are three types of channels in an IRS-aided
system: the \textbf{direct channel} from the BS to the user, the \textbf{indirect
channel} from the BS to the IRS and the \textbf{reflection channel}
from the IRS to the user. The first one can be obtained with high
accuracy by using conventional channel estimation methods. The accurate
CSI of the latter two, however, are challenging to obtain in practice
due to the fact that the reflective elements at the IRS are passive
and have limited signal processing capability. Fortunately, the location
of the IRS is fixed and is usually installed in the building facades,
ceilings, walls, etc. In this case, the \textbf{indirect channel}
can be accurately estimated through calculating the angles of arrival
and departure, which vary slowly. In contrast, the \textbf{reflection
channel} is more challenging to acquire as the locations of users
are changing and their environmental conditions are varying.

Against the above background, this paper investigates the robust active
precoder and passive reflection beamforming design for an IRS-aided
downlink multiple-user multiple-input single-output (MU-MISO) system
based on the assumption of imperfect \textbf{reflection channel}.
An ellipsoid model of the reflection channel uncertainties are adopted.
To the best of our knowledge, this is the first work to study the
worst-case robust beamforming design problem in IRS-aided wireless
systems. The contributions of this paper are as follows:
1) We aim to minimize the transmit power of the BS through the joint
design of an active precoder at the BS and a passive beamforming at
the IRS while ensuring that each user's QoS target can be achieved
for all possible channel error realizations. This problem is non-convex
and difficult to solve due to the unit-modulus constraints and the
imperfect CSI. 2) To address this problem, we propose an iterative
algorithm based on approximation transformations and a convex--concave
procedure (CCP). Specifically, to handle the non-convex rate expression
and CSI uncertainties, we first approximately linearize the rates
by using the first-order Taylor expansion, and then transform the
resultant semi-infinite constraints into linear matrix inequalities
(LMIs). The non-convex unit-modulus constraints of the reflection
beamforming are handled by the penalized CCP \cite{PCCP-boyd}. 3)
Numerical results confirm the effectiveness of the proposed algorithms
in guaranteeing the QoS targets of all users.

\section{System Model}

\subsection{Signal Transmission Model}

\begin{figure}
\centering \includegraphics[width=2.8in,height=1.6in]{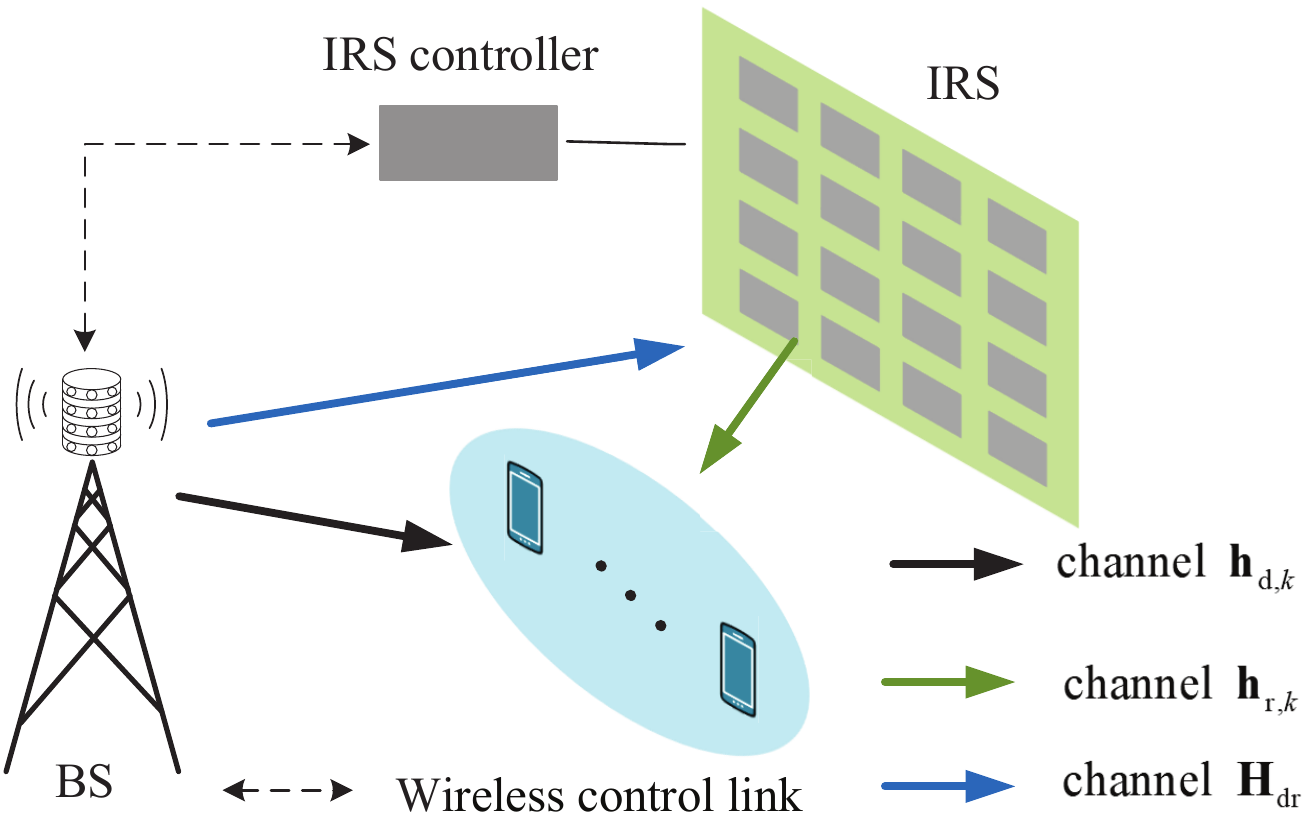}
\caption{An IRS-aided multiuser communication system.}
\label{system-model} 
\end{figure}

We consider an IRS-aided MISO broadcast (BC) communication system
shown in Fig. \ref{system-model}, in which there is a BS equipped
with $N$ transmit antennas serving $K$ single-antenna users. Denote
by $\mathbf{s}=[s_{1},\cdots,s_{K}]^{\mathrm{T}}\in\mathbb{C}^{K\times1}$
the Gaussian data symbols, in which each element is an independent
random variable with zero mean and unit variance, i.e., $\mathbb{E}[\mathbf{s}\mathbf{s}^{\mathrm{H}}]=\mathbf{I}$.
Denote by $\mathbf{F=}[{\bf \mathbf{f}}_{1},\cdots,{\bf \mathbf{f}}_{K}]\in\mathbb{C}^{N\times K}$
the corresponding precoding vectors for the users. Then, the transmit
signal at the BS is ${\bf x}={\bf F}{\bf s}$, the transmit power
of which is $\mathbb{E}\{\mathrm{Tr}\left[\mathbf{x}\mathbf{x}^{\mathrm{H}}\right]\}=||\mathbf{F}||_{F}^{2}$.

In the MISO BC system, we propose to employ an IRS with the goal of
enhancing the received signal strength of the users by reflecting
signals from the BS to the users. It is assumed that the IRS has $M$
passive reflection elements $\mathbf{e}=[e_{1},\cdots,e_{M}]^{\mathrm{T}}\in\mathbb{C}^{M\times1}$,
the modulus of each element is $|e_{m}|^{2}=1,1\leq m\leq M$. Then,
the reflection beamforming at the IRS is modeled as a diagonal matrix
$\mathbf{E}=\iota\mathrm{diag}(\mathbf{e})\in\mathbb{C}^{M\times M}$
where $\iota\in[0,1]$ indicates the reflection efficiency. The channels
from the BS to user $k$, from the BS to the IRS, and from the IRS
to user $k$ are denoted by $\mathbf{h}_{\mathrm{d},k}\in\mathbb{C}^{N\times1}$,
$\mathbf{H_{\mathrm{dr}}}\in\mathbb{C}^{M\times N}$, and $\mathbf{h}_{\mathrm{r},k}\in\mathbb{C}^{M\times1}$,
respectively.

The BS is responsible for designing the reflection beamforming at
the IRS and sending it to the IRS controller \cite{Pan2019intelleget}.
Let us define the set of all users as $\mathcal{K}=\{1,2,...,K\}$
, then the received signal of the users is 
\begin{equation}
y_{k}=(\mathbf{h}_{\mathrm{d},k}^{\mathrm{H}}+\mathbf{h}_{\mathrm{r},k}^{\mathrm{H}}\mathbf{E}\mathbf{H_{\mathrm{dr}}}){\bf F}{\bf s}+n_{k},\forall k\in\mathcal{K},\label{eq:received signal}
\end{equation}
where $n_{k}$ is the received noise at user $k$, which is an additive
white Gaussian noise (AWGN) with distribution $\mathcal{CN}(0,\sigma_{k}^{2})$.
The achievable data rate (bit/s/Hz) at user $k$ is given by 
\begin{align}
R_{k}\left(\mathbf{F},\mathbf{e}\right)= & \log_{2}\left(1+\left|\left(\mathbf{h}_{\mathrm{d},k}^{\mathrm{H}}+\mathbf{h}_{\mathrm{r},k}^{\mathrm{H}}\mathbf{E}\mathbf{H_{\mathrm{dr}}}\right){\bf f}_{k}\right|^{2}/\beta_{k}\right)\label{eq:eq:rate-1}
\end{align}
where $\beta_{k}=||(\mathbf{h}_{\mathrm{d},k}^{\mathrm{H}}+\mathbf{h}_{\mathrm{r},k}^{\mathrm{H}}\mathbf{E}\mathbf{H_{\mathrm{dr}}}){\bf F}_{-k}||_{2}^{2}+\sigma_{k}^{2},\forall k\in\mathcal{K}$
represent the interference-plus-noises (INs) term with ${\bf F}_{-k}=[{\bf \mathbf{f}}_{1},\cdots,{\bf \mathbf{f}}_{k-1},{\bf \mathbf{f}}_{k+1},\cdots,{\bf \mathbf{f}}_{K}]$.

In the IRS-aided communication system, there are three types of channels:
the direct channel from the BS to the user, i.e., $\mathbf{h}_{\mathrm{d},k}$,
the indirect channel from the BS to the IRS, i.e., $\mathbf{H_{\mathrm{dr}}}$,
and the reflection channel from the IRS to the user, i.e., $\mathbf{h}_{\mathrm{r},k}$.
As mentioned in the introduction section, the reflection channel is
much more challenging to obtain than the other two channels. Hence,
in this paper, we assume that the third type of channel is imperfect.
The reflection channel $\{\mathbf{h}_{\mathrm{r},k}\}_{\forall k\in\mathcal{K}}$
can be modeled as $\{\mathbf{h}_{\mathrm{r},k}=\widehat{\mathbf{h}}_{\mathrm{r},k}+\boldsymbol{\triangle}_{k}\}_{\forall k\in\mathcal{K}}$,
where $\{\widehat{\mathbf{h}}_{\mathrm{r},k}\}_{\forall k\in\mathcal{K}}$
denote the contaminated channel vectors and $\{\boldsymbol{\triangle}_{k}\}_{\forall k\in\mathcal{K}}$
denote the corresponding channel error vectors. In this paper, we
adopt the channel error bounded model, i.e., $\{||\boldsymbol{\triangle}_{k}||_{2}\leq\varepsilon_{k}\}_{\forall k\in\mathcal{K}}$,
where $\varepsilon_{k}$ is the radius of the uncertainty region known
by the BS.

\vspace{-0.4cm}

\subsection{Problem Formulation}

With imperfect CSI, we aim to minimize the total transmit power via
the joint design of the precoding matrix $\mathbf{F}$ and the reflection
vector $\mathbf{e}$ under the worst-case QoS constraints, i.e., ensuring
that the achievable rate of each user is above a threshold for all
possible channel error realizations. Mathematically, the worst-case
robust design problem is formulated as \begin{subequations}\label{Pro:min-power}
\begin{align}
\mathop{\min}\limits _{\mathbf{F},\mathbf{e}} & \;\;||\mathbf{F}||_{F}^{2}\label{eq:min-power-obj}\\
\textrm{s.t.} & \thinspace\thinspace\thinspace R_{k}\left(\mathbf{F},\mathbf{e}\right)\geq r_{k},\forall\left\Vert \boldsymbol{\triangle}_{k}\right\Vert _{2}\leq\varepsilon_{k},\forall k\in\mathcal{K},\label{eq:min-power-cons1}\\
 & \thinspace\thinspace\thinspace|e_{m}|^{2}=1,1\leq m\leq M.\label{eq:min-power-cons2}
\end{align}
\end{subequations}Constraints (\ref{eq:min-power-cons1}) are the
minimum QoS targets for each user, while constraints (\ref{eq:min-power-cons2})
correspond to the unit-modulus requirements of the reflection elements
at the IRS.

\vspace{-0.4cm}

\section{Robust beamforming design}

Problem (\ref{Pro:min-power}) is a non-convex problem and the main
challenge lies in the non-convex QoS constraints (\ref{eq:min-power-cons1})
over the CSI uncertainty regions and the non-convex unit-modulus constraints
(\ref{eq:min-power-cons2}). Since variables $\mathbf{F}$ and $\mathbf{e}$
are coupled, we propose an alternate optimization (AO) method to solve
Problem (\ref{Pro:min-power}).

\vspace{-0.4cm}

\subsection{Problem Transformation}

To start with, the non-convexity of constraints (\ref{eq:min-power-cons1})
can be addressed by firstly treating the INs $\boldsymbol{\beta}=[\beta_{1},...,\beta_{K}]^{\mathrm{T}}$
as auxiliary variables. Hence, constraints (\ref{eq:min-power-cons1})
are rewritten as \begin{subequations}\label{Pro:min-power-1} 
\begin{align}
 & \left|\left(\mathbf{h}_{\mathrm{d},k}^{\mathrm{H}}+\mathbf{h}_{\mathrm{r},k}^{\mathrm{H}}\mathbf{E}\mathbf{H_{\mathrm{dr}}}\right){\bf f}_{k}\right|^{2}\geq\beta_{k}(2^{r_{k}}-1),\nonumber \\
 & \forall\left\Vert \boldsymbol{\triangle}_{k}\right\Vert _{2}\leq\varepsilon_{k},\forall k\in\mathcal{K},\label{eq:single}\\
 & \left\Vert \left(\mathbf{h}_{\mathrm{d},k}^{\mathrm{H}}+\mathbf{h}_{\mathrm{r},k}^{\mathrm{H}}\mathbf{E}\mathbf{H_{\mathrm{dr}}}\right){\bf F}_{-k}\right\Vert _{2}^{2}+\sigma_{k}^{2}\leq\beta_{k},\nonumber \\
 & \forall\left\Vert \boldsymbol{\triangle}_{k}\right\Vert _{2}\leq\varepsilon_{k},\forall k\in\mathcal{K}.\label{eq:IN}
\end{align}
\end{subequations}

We first handle the infinite inequalities in (\ref{eq:single}), which
are non-convex. Specifically, the left hand side (LHS) of (\ref{eq:single})
is approximated as its lower bound, as shown below..

\vspace{-0.4cm}
\begin{lemma}\label{lower-bound} Let $\mathbf{f}_{k}^{(n)}$ and
$\mathbf{E}^{(n)}$ be the optimal solutions obtained at iteration
$n$, then the linear lower bound of $|(\mathbf{h}_{\mathrm{d},k}^{\mathrm{H}}+\mathbf{h}_{\mathrm{r},k}^{\mathrm{H}}\mathbf{E}\mathbf{H_{\mathrm{dr}}}){\bf f}_{k}|^{2}$
in (\ref{eq:single}) at ($\mathbf{f}_{k}^{(n)}$, $\mathbf{E}^{(n)}$)
is 
\begin{equation}
\mathbf{h}_{\mathrm{r},k}^{\mathrm{H}}\mathbf{X}_{k}\mathbf{h}_{\mathrm{r},k}+\mathbf{h}_{\mathrm{r},k}^{\mathrm{H}}\mathbf{x}_{k}+\mathbf{x}_{k}^{\mathrm{H}}\mathbf{h}_{\mathrm{r},k}+c_{k},\label{eq:lower-bound-1}
\end{equation}
where
\begin{align*}
\mathbf{X}_{k} & =\mathbf{E}\mathbf{H_{\mathrm{dr}}}{\bf f}_{k}{\bf f}_{k}^{\mathrm{H},(n)}\mathbf{H_{\mathrm{dr}}^{\mathrm{H}}}\mathbf{E}^{\mathrm{H},(n)}+\mathbf{E}^{(n)}\mathbf{H_{\mathrm{dr}}}{\bf f}_{k}^{(n)}{\bf f}_{k}^{\mathrm{H}}\mathbf{H_{\mathrm{dr}}^{\mathrm{H}}}\mathbf{E}^{\mathrm{H}}\\
 & \thinspace\thinspace\thinspace\thinspace\thinspace-\mathbf{E}^{(n)}\mathbf{H_{\mathrm{dr}}}{\bf f}_{k}^{(n)}{\bf f}_{k}^{\mathrm{H},(n)}\mathbf{H_{\mathrm{dr}}^{\mathrm{H}}}\mathbf{E}^{\mathrm{H},(n)},\\
\mathbf{x}_{k} & =\mathbf{E}\mathbf{H_{\mathrm{dr}}}{\bf f}_{k}{\bf f}_{k}^{\mathrm{H},(n)}\mathbf{h}_{\mathrm{d},k}+\mathbf{E}^{(n)}\mathbf{H_{\mathrm{dr}}}{\bf f}_{k}^{(n)}{\bf f}_{k}^{\mathrm{H}}\mathbf{h}_{\mathrm{d},k}\\
 & \thinspace\thinspace\thinspace\thinspace\thinspace-\mathbf{E}^{(n)}\mathbf{H_{\mathrm{dr}}}{\bf f}_{k}^{(n)}{\bf f}_{k}^{\mathrm{H},(n)}\mathbf{h}_{\mathrm{d},k},\\
c_{k} & =\mathbf{h}_{\mathrm{d},k}^{\mathrm{H}}({\bf f}_{k}{\bf f}_{k}^{\mathrm{H},(n)}+{\bf f}_{k}^{(n)}{\bf f}_{k}^{\mathrm{H}}-{\bf f}_{k}^{(n)}{\bf f}_{k}^{\mathrm{H},(n)})\mathbf{h}_{\mathrm{d},k}.
\end{align*}
\end{lemma}

$\mathbf{Proof:}$ Let $a$ be a complex scalar variable. By applying
Appendix B of \cite{Pan2019robust}, we have the inequality 
\begin{equation}
\left|a\right|^{2}\geq a^{*,(n)}a+a^{*}a^{(n)}-a^{*,(n)}a^{(n)}\label{eq:df}
\end{equation}
for any fixed $a^{(n)}$. Then, (\ref{eq:lower-bound-1}) is obtained
by replacing $a$ and $a^{(n)}$ with $(\mathbf{h}_{\mathrm{d},k}^{\mathrm{H}}+\mathbf{h}_{\mathrm{r},k}^{\mathrm{H}}\mathbf{E}\mathbf{H_{\mathrm{dr}}}){\bf f}_{k}$
and $(\mathbf{h}_{\mathrm{d},k}^{\mathrm{H}}+\mathbf{h}_{\mathrm{r},k}^{\mathrm{H}}\mathbf{E}^{(n)}\mathbf{H_{\mathrm{dr}}}){\bf f}_{k}^{(n)}$,
respectively. The proof is complete. \hspace{0.1cm}$\blacksquare$

With $\mathbf{h}_{\mathrm{r},k}=\widehat{\mathbf{h}}_{\mathrm{r},k}+\boldsymbol{\triangle}_{k}$
and Lemma \ref{lower-bound}, the inequality (\ref{eq:single}) is
reformulated as
\begin{align}
 & \triangle\mathbf{h}_{\mathrm{r},k}^{\mathrm{H}}\mathbf{X}_{k}\triangle\mathbf{h}_{\mathrm{r},k}+2\mathrm{Re}\left\{ (\mathbf{x}_{k}^{\mathrm{H}}+\widehat{\mathbf{h}}_{\mathrm{r},k}^{\mathrm{H}}\mathbf{X}_{k})\triangle\mathbf{h}_{\mathrm{r},k}\right\} +d_{k}\nonumber \\
 & \geq\beta_{k}(2^{r_{k}}-1),\forall\left\Vert \boldsymbol{\triangle}_{k}\right\Vert _{2}\leq\varepsilon_{k},\forall k\in\mathcal{K},\label{eq:signal-2}
\end{align}
where $d_{k}=\widehat{\mathbf{h}}_{\mathrm{r},k}^{\mathrm{H}}\mathbf{X}_{k}\widehat{\mathbf{h}}_{\mathrm{r},k}+\mathbf{x}_{k}^{\mathrm{H}}\widehat{\mathbf{h}}_{\mathrm{r},k}+\widehat{\mathbf{h}}_{\mathrm{r},k}^{\mathrm{H}}\mathbf{x}_{k}+c_{k}$.

In order to tackle the CSI uncertainties, the S-Procedure in \cite{Luo2004SIAM}
is used to transform (\ref{eq:signal-2}) into equivalent LMIs as
\begin{align}
 & \left[\begin{array}{cc}
\varpi_{k}\mathbf{I}_{M}+\mathbf{X}_{k} & (\mathbf{x}_{k}^{\mathrm{H}}+\widehat{\mathbf{h}}_{\mathrm{r},k}^{\mathrm{H}}\mathbf{X}_{k})^{\mathrm{H}}\\
(\mathbf{x}_{k}^{\mathrm{H}}+\widehat{\mathbf{h}}_{\mathrm{r},k}^{\mathrm{H}}\mathbf{X}_{k}) & d_{k}-\beta_{k}(2^{r_{k}}-1)-\varpi_{k}\varepsilon_{k}^{2}
\end{array}\right]\succeq\mathbf{0},\nonumber \\
 & \forall k\in\mathcal{K},\label{eq:LMI-signal}
\end{align}
where $\boldsymbol{\varpi}=[\varpi_{1},...,\varpi_{K}]^{\mathrm{T}}\geq0$
are slack variables.

Now, we consider the uncertainties in $\{\boldsymbol{\triangle}_{k}\}_{\forall k\in\mathcal{K}}$
of (\ref{eq:IN}). To this end, we first adopt Schur's complement
\cite{book-convex} to equivalently recast (\ref{eq:IN}) as 
\begin{align}
 & \left[\begin{array}{cc}
\beta_{k}-\sigma_{k}^{2} & \mathbf{t}_{k}^{\mathrm{H}}\\
\mathbf{t}_{k} & \mathbf{I}
\end{array}\right]\succeq\mathbf{0},\forall\left\Vert \boldsymbol{\triangle}_{k}\right\Vert _{2}\leq\varepsilon_{k},\forall k\in\mathcal{K},\label{eq:IN-LMI-1}
\end{align}
where $\mathbf{t}_{k}=((\mathbf{h}_{\mathrm{d},k}^{\mathrm{H}}+\mathbf{h}_{\mathrm{r},k}^{\mathrm{H}}\mathbf{E}\mathbf{H_{\mathrm{dr}}}){\bf F}_{-k})^{\mathrm{H}}$.

Then, by using Nemirovski lemma \cite{Ben-Tal2005robust} and introducing
the slack variables $\boldsymbol{\xi}=[\xi_{1},...,\xi_{K}]^{\mathrm{T}}\geq0$,
(\ref{eq:IN-LMI-1}) is rewritten as 
\begin{align}
 & \left[\begin{array}{ccc}
\beta_{k}-\sigma_{k}^{2}-\xi_{k} & \widehat{\mathbf{t}}_{k}^{\mathrm{H}} & \mathbf{0}_{1\times M}\\
\widehat{\mathbf{t}}_{k} & \mathbf{I}_{(K-1)} & \varepsilon_{k}\left(\mathbf{E}\mathbf{H_{\mathrm{dr}}}{\bf F}_{-k}\right)^{\mathrm{H}}\\
\mathbf{0}_{M\times1} & \varepsilon_{k}\mathbf{E}\mathbf{H_{\mathrm{dr}}}{\bf F}_{-k} & \xi_{k}\mathbf{I}_{M}
\end{array}\right]\succeq\mathbf{0},\nonumber \\
 & \forall k\in\mathcal{K},\label{eq:LMI-IN}
\end{align}
where $\widehat{\mathbf{t}}_{k}=((\mathbf{h}_{\mathrm{d},k}^{\mathrm{H}}+\widehat{\mathbf{h}}_{\mathrm{r},k}^{\mathrm{H}}\mathbf{E}\mathbf{H_{\mathrm{dr}}}){\bf F}_{-k})^{\mathrm{H}}$.

With (\ref{eq:LMI-signal}) and (\ref{eq:LMI-IN}), we obtain the
following approximated reformulation of Problem (\ref{Pro:min-power})
as \begin{subequations}\label{Pro:min-power-2} 
\begin{align}
\mathop{\min}\limits _{\mathbf{F},\mathbf{e},\boldsymbol{\beta},\boldsymbol{\varpi},\boldsymbol{\xi}} & \;\;||\mathbf{F}||_{F}^{2}\label{eq:obj-2}\\
{\rm s.t.} & \thinspace\thinspace\thinspace(\ref{eq:LMI-signal}),(\ref{eq:LMI-IN}),(\ref{eq:min-power-cons2}),\\
 & \thinspace\thinspace\thinspace\boldsymbol{\varpi}\geq0,\boldsymbol{\xi}\geq0.\label{eq:omiga}
\end{align}
\end{subequations}

It is difficult to optimize the variables $\mathbf{F}$ and $\mathbf{e}$
simultaneously as they are coupled in the LMIs (\ref{eq:LMI-signal})
and (\ref{eq:LMI-IN}). Therefore, the AO method is adopted to solve
the subproblems corresponding to different sets of variables iteratively.
Specifically, for a given reflection beamforming $\mathbf{e}$, the
subproblem of Problem (\ref{Pro:min-power-2}) corresponding to the
precoder $\mathbf{F}$ is formulated as \begin{subequations}\label{Pro:min-power-3}
\begin{align}
\mathbf{F}^{(n+1)}=\mathrm{arg}\mathop{\min}\limits _{\mathbf{F},\boldsymbol{\beta},\boldsymbol{\varpi},\boldsymbol{\xi}} & \;\;||\mathbf{F}||_{F}^{2}\label{eq:obj-3}\\
{\rm s.t.} & \thinspace\thinspace\thinspace(\ref{eq:LMI-signal}),(\ref{eq:LMI-IN}),(\ref{eq:omiga})\text{,}
\end{align}
\end{subequations} where $\mathbf{F}^{(n+1)}$ is the optimal solution
obtained in the $(n+1)$-th iteration. Problem (\ref{Pro:min-power-3})
is a semidefinite program (SDP) and can be solved by using the CVX
tool.

On the other hand, for a given precoding matrix $\mathbf{F}$, the
subproblem of Problem (\ref{Pro:min-power-2}) corresponding to $\mathbf{e}$
is a feasibility-check problem. According to the Problem (P4') in
\cite{qingqing2019}, the converged solution in the optimization of
$\mathbf{e}$ can be improved by introducing slack variables $\mathbf{a}=[a_{1},...,a_{K}]^{\mathrm{T}}$
which are interpreted as the ``signal-to-interference-plus-noise
ratio (SINR) residual'' of users. Please refer to
\cite{qingqing2019} about the theory of ``SINR residual''. Thus,
the feasibility-check problem of $\mathbf{e}$ is formulated as follows
\begin{subequations}\label{Pro:min-power-4} 
\begin{align}
\max\limits _{\mathbf{e},\mathbf{a},\boldsymbol{\beta},\boldsymbol{\varpi},\boldsymbol{\xi}} & \;||\mathbf{a}||_{1}\label{eq:obj-4}\\
{\rm s.t.} & \thinspace\thinspace\thinspace\textrm{Modified-}(\ref{eq:LMI-signal}),(\ref{eq:LMI-IN}),(\ref{eq:min-power-cons2}),(\ref{eq:omiga}),\\
 & \thinspace\thinspace\thinspace\mathbf{a}\geq0,\label{eq:SINR residual}
\end{align}
\end{subequations}where the Modified-(\ref{eq:LMI-signal}) constraints
are LMIs obtained from (\ref{eq:LMI-signal}) by replacing $\beta_{k}(2^{r_{k}}-1)$
with $\beta_{k}(2^{r_{k}}-1)+a_{k}$ for $\forall k\in\mathcal{K}$.

However, the above problem cannot be solved directly due to the non-convex
constraint (\ref{eq:min-power-cons2}). In addition, the semidefinite
relaxation (SDR) method used in \cite{qingqing2019} cannot always
guarantee a feasible solution due to the fact that the QoS constraints
may be violated when the SDR solution is not rank one. To handle this
issue, we apply the penalty CCP \cite{PCCP-boyd} which is capable
of finding a feasible solution that meets the unit-modulus constraint
and the QoS constraints. In particular, the constraints $|e_{m}|^{2}=1,1\leq m\leq M$
can be equivalently rewritten as $1\leq|e_{m}|^{2}\leq1,1\leq m\leq M$.
The non-convex parts of these constraints are again linearized by
$|e_{m}^{{\text{[}t]}}|^{2}-2\mathrm{Re}(e_{m}^{\mathrm{H}}e_{m}^{{[t]}})\leq-1,1\leq m\leq M$
at fixed $e_{m}^{{[t]}}$. Following the penalty CCP framework,
we impose the use of slack variables $\mathbf{b}=[b_{1},...,b_{2M}]^{\mathrm{T}}$
over the equivalent constraints of the unit-modulus constraints, which
yields\begin{subequations}\label{Pro:min-power-5} 
\begin{align}
\max\limits _{{\scriptstyle {\mathbf{e},\mathbf{a},\mathbf{b},\hfill\atop {\scriptstyle \boldsymbol{\beta},\boldsymbol{\varpi},\boldsymbol{\xi}\hfill}}}} & \;||\mathbf{a}||_{1}-\lambda^{{[t]}}||\mathbf{b}||_{1}\label{eq:obj-4-1}\\
{\rm s.t.} & \textrm{\thinspace\thinspace\,Modified-(\ref{eq:LMI-signal}),(\ref{eq:LMI-IN}),(\ref{eq:omiga}),(\ref{eq:SINR residual})},\\
 & \thinspace|e_{m}^{{\text{[}t]}}|^{2}-2\mathrm{Re}(e_{m}^{\mathrm{H}}e_{m}^{{[t]}})\leq b_{m}-1,1\leq m\leq M\label{eq:unit-3-1}\\
 & \thinspace\thinspace\thinspace|e_{m}|^{2}\leq1+b_{M+m},1\leq m\leq M\\
 & \thinspace\thinspace\thinspace\mathbf{b}\geq0,
\end{align}
\end{subequations}where $\lambda^{{[t]}}$ is the regularization
factor to scale the impact of the penalty term $||\mathbf{b}||_{1}$,
which controls the feasibility of the constraints. At low $\lambda$,
Problem (\ref{Pro:min-power-5}) targets to maximize the ``SINR residual'',
while Problem (\ref{Pro:min-power-5}) seeks for a feasible point
rather than optimizing the ``SINR residual'' at high $\lambda$.

{Problem (\ref{Pro:min-power-5}) is an SDP and can
be solved by using the CVX tool.} The algorithm for finding a feasible
solution of $\mathbf{e}$ is summarized in Algorithm \ref{Algorithm-analog}.
Some points are emphasized as follows: \textit{a) }The maximum value
$\lambda_{max}$ is imposed to avoid numerical problems, that is,
a feasible solution may not be found when the iteration converges
under increasing large values of $\lambda^{[t]}$;\textit{
b) }The stopping criteria $||\mathbf{b}||_{1}\leq\chi$ guarantees
the unit-modulus constraints in the original Problem (\ref{Pro:min-power-4})
to be met for a sufficiently low $\chi$; \textit{c) }The stopping
criteria $||\mathbf{e}^{{[t]}}-\mathbf{e}^{{[t-1]}}||_{1}\leq\nu$
controls the convergence of Algorithm \ref{Algorithm-analog}; \textit{d)}
As mentioned in \cite{PCCP-boyd}, a feasible solution for Problem
(\ref{Pro:min-power-5}) may not be feasible for Problem (\ref{Pro:min-power-4}).
Hence, the feasibility of Problem (\ref{Pro:min-power-4}) is guaranteed
by imposing a maximum number of iterations $T_{max}$ and, in case
it is reached, we restart the iteration based on a new initial point.

\begin{algorithm}
\caption{Penalty CCP optimization for reflection beamforming optimization}
\label{Algorithm-analog} \begin{algorithmic}[1] \REQUIRE Initialize
$\mathbf{e}^{{[0]}}$, $\gamma^{{[0]}}>1$,
and set $t=0$.

\REPEAT

\IF {$t<T_{max}$ }

\STATE Update $\mathbf{e}^{{[t+1]}}$ from Problem (\ref{Pro:min-power-5});

\STATE $\lambda^{{[t+1]}}=\min\{\gamma\lambda^{{[t]}},\lambda_{max}\}$;

\STATE $t=t+1$;

\ELSE

\STATE Initialize with a new random $\mathbf{e}^{[0]}$,
set $\gamma^{[0]}>1$, and $t=0$.

\ENDIF

\UNTIL $||\mathbf{b}||_{1}\leq\chi$ and $||\mathbf{e}^{[t]}-\mathbf{e}^{[t-1]}||_{1}\leq\nu$.

\STATE Output $\mathbf{e}^{(n+1)}=\mathbf{e}^{[t]}$.

\end{algorithmic} 
\end{algorithm}

\vspace{-0.5cm}

\subsection{Algorithm Description}

Algorithm \ref{Algorithm-AO} summarizes the AO method for solving
Problem (\ref{Pro:min-power-2}).

\begin{algorithm}
\caption{AO algorithm for Problem (\ref{Pro:min-power-2})}
\label{Algorithm-AO} \begin{algorithmic}[1] \REQUIRE Initialize
$\mathbf{e}^{(0)}$ and $\mathbf{F}^{(0)}$, and set $n=0$.

\REPEAT

\STATE Update $\mathbf{F}^{(n+1)}$ from Problem (\ref{Pro:min-power-3})
with given $\mathbf{e}^{(n)}$;

\STATE Update $\mathbf{e}^{(n+1)}$ from Problem (\ref{Pro:min-power-4})
with given $\mathbf{F}^{(n+1)}$;

\STATE $n\leftarrow n+1$;

\UNTIL The objective value $||\mathbf{F}^{(n+1)}||_{F}^{2}$ converges.

\end{algorithmic} 
\end{algorithm}

\paragraph{Convergence analysis}

The convergence of Algorithm \ref{Algorithm-AO} can be guaranteed.
In particular, denoting the objective value of Problem (\ref{Pro:min-power-3})
as $F(\mathbf{F},\mathbf{e})$, it follows that 
\begin{align*}
 & F(\mathbf{F}^{(n)},\mathbf{e}^{(n)})\geq F(\mathbf{F}^{(n)},\mathbf{e}^{(n+1)})\geq F(\mathbf{F}^{(n+1)},\mathbf{e}^{(n+1)}).
\end{align*}
The above equality holds true because the objective value of Problem
(\ref{Pro:min-power-3}) is independent of $\mathbf{e}$, and also
$\mathbf{e}^{(n+1)}$ is feasible for Problem (\ref{Pro:min-power-3})
if it is a feasible solution for Problem (\ref{Pro:min-power-4}).
The above inequality follows from the globally optimal solution $\mathbf{F}^{(n+1)}$
of Problem (\ref{Pro:min-power-3}) for a given $\mathbf{e}^{(n+1)}$.
Hence, the sequence $\{F(\mathbf{F}^{(n)},\mathbf{e}^{(n)})\}$ is
non-increasing and the algorithm is guaranteed to converge.

\paragraph{Initial point}

As for the method of initialization, $\mathbf{e}^{(0)}$ can be chosen
as a full-1 vector for simplicity. Inspired by \cite{pan2017JSAC},
the initial point $\mathbf{F}^{(0)}$ can be chosen as the optimal
solution to the following optimization problem \begin{subequations}\label{Pro:min-power-6}
\begin{align}
\mathop{\min}\limits _{\mathbf{F},\boldsymbol{\varphi}} & \;\;\sum_{k\in\mathcal{K}}(\varphi_{k}-1)^{2}\label{eq:min-power-obj-1}\\
\textrm{s.t.} & \thinspace\thinspace\thinspace R_{k}\left(\mathbf{F},\mathbf{e}\right)\geq\varphi_{k}\mathrm{r_{k}},\forall k\in\mathcal{K}\label{eq:SDF}\\
 & \varphi_{k}\geq0,\forall k\in\mathcal{K},
\end{align}
\end{subequations}where $\boldsymbol{\varphi}=[\varphi_{1},...,\varphi_{K}]^{\mathrm{T}}$
is an auxiliary variable vector. Problem (\ref{Pro:min-power-6})
is guaranteed to be feasible since at least $\{\varphi_{k}=0,\forall k\in\mathcal{K},\mathbf{F}^{(n)}=\mathbf{0}\}$
is a feasible solution. Problem (\ref{Pro:min-power-6}) can also
be solved by reformulating it into an alternative optimization problem
that is similar to Problems (\ref{Pro:min-power-3}). Denote by $\{\varphi_{k}^{(\mathrm{opt})}\}_{\forall k\in\mathcal{K}}$
the solution of Problem (\ref{Pro:min-power-6}), then the corresponding
optimal precoding matrix can be used as the initial point for Algorithm
\ref{Algorithm-AO} if $\varphi_{k}^{(\mathrm{opt})}=1,\forall k\in\mathcal{K}$.

\section{Numerical results and discussions}

In this section, numerical results are provided to evaluate the performance
of the proposed algorithm. We consider that the BS is equipped with
$N=6$ transmit antennas serving $K=4$ users with
the assistance of an IRS. The number of the reflection elements is
$M=16$. We assume a rectangular coordinate to discribe
the system, i.e., the locations of the BS and IRS are (0 m, 0 m) and
(50 m, 10 m) respectively, and users are distributed randomly on a
circle centered at (70 m, 0 m) with radius 5 m.

The large-scale path loss is $\mathrm{PL}=-30-10\alpha\log_{10}(d)$
dB, where $\alpha$ is the path loss exponent and $d$ is the link
length in meters. The path loss exponents for the BS-IRS link, BS-user
link, and the IRS-user link are equal to $\alpha_{\mathrm{BI}}=2.2$
\cite{Marco-1}, $\alpha_{\mathrm{BU}}=4$ and $\alpha_{\mathrm{IU}}=2$,
respectively. The small-scale fading of the channels $[\mathbf{H_{\mathrm{dr}}},\{\mathbf{h}_{\mathrm{d},k},\widehat{\mathbf{h}}_{\mathrm{r},k}\}_{\forall k\in\mathcal{K}}]$
follows a Rician distribution with Ricean factor 5.
The line-of-sight (LoS) components are defined by the product of the
steering vectors of the transmitter and receiver and the non-LoS components
are drawn from a Rayleigh fading. The CSI error bounds are defined
as $\varepsilon_{k}=\delta||\widehat{\mathbf{h}}_{\mathrm{r},k}||_{2},\forall k\in\mathcal{K}$,
where $\delta\in[0,1)$ accounts for the relative amount of CSI uncertainties.
The power of the AWGN at all users is set to $-100$
dBm and the target rates of all users are the same, i.e., $r_{1}=...=r_{K}=r$.
The IRS and benchmark schemes considered are the following: 1) ``IRS,
$\iota=1$(or 0.5)''. 2) ``Non-robust IRS'', in which the channel
estimation error is ignored when designing the beamformings. 3) ``Non
IRS'', in which there is no IRS in the MU-MISO system. 4) ``Relay'',
in which a full-duplex relay is located at the same place of the IRS.
The numbers of transmit and receive antennas at the relay are both
$M$.

\begin{figure}
\centering \subfigure[The transmit power]{ %
\begin{minipage}[t]{0.495\linewidth}%
\centering \includegraphics[width=1.7in]{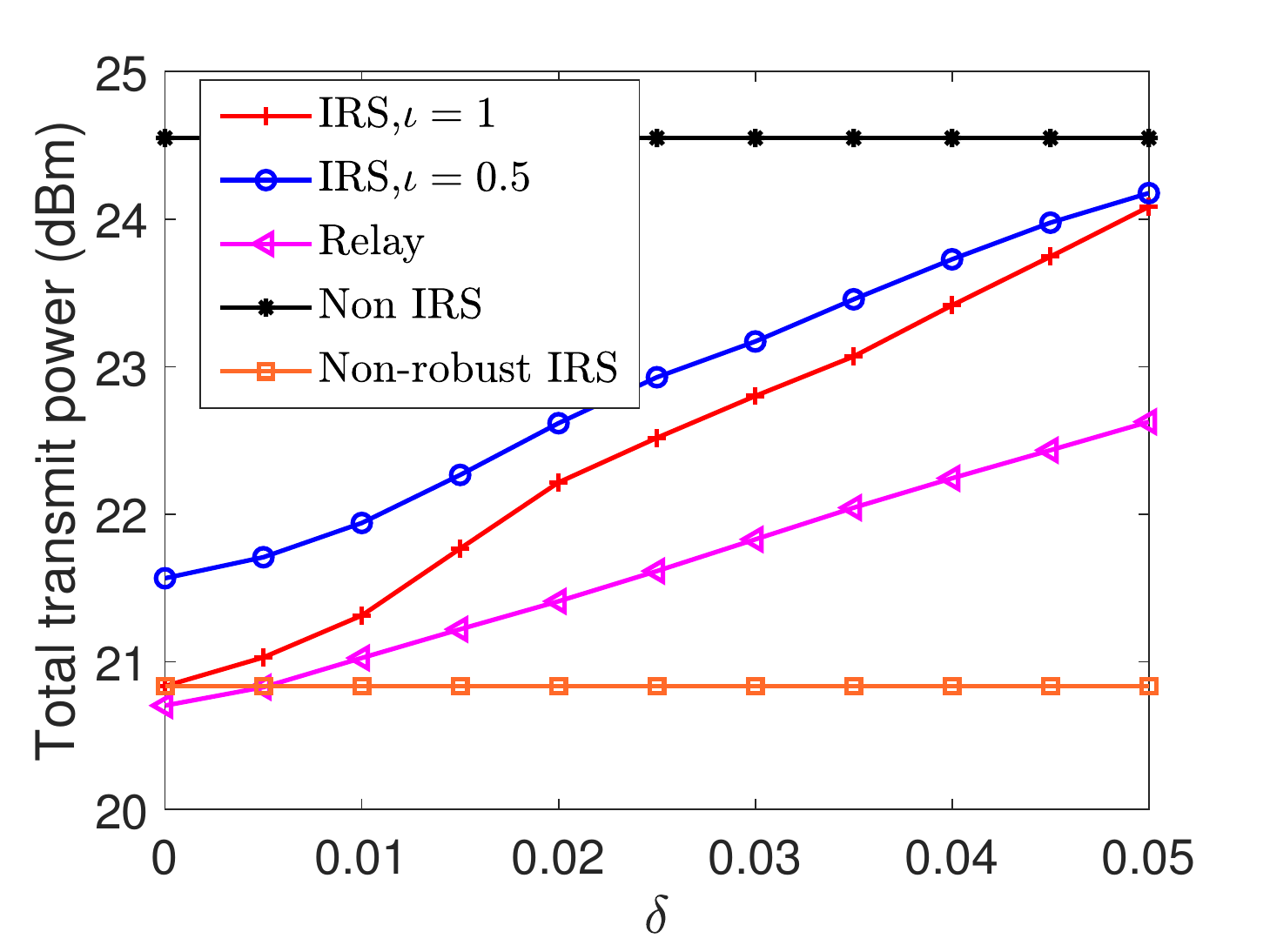} 
\end{minipage}}\subfigure[The energy efficiency]{ %
\begin{minipage}[t]{0.495\linewidth}%
\centering \includegraphics[width=1.7in]{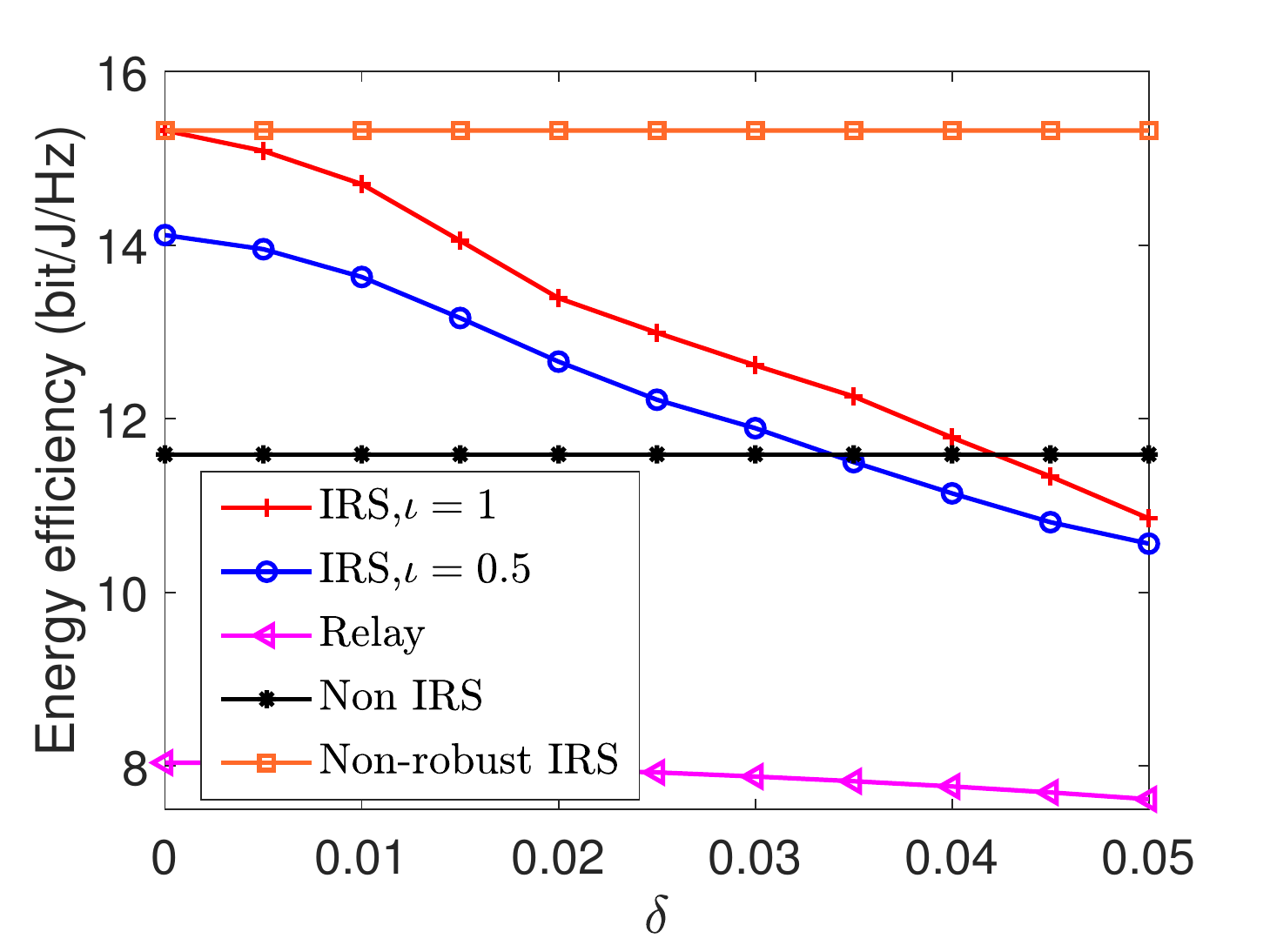} 
\end{minipage}}

\caption{Performance versus the channel uncertainty level $\delta$
under $N=6$, $M=16$, $K=4$ and $r=4$ bit/s/Hz.}
\label{error}
\end{figure}

Firstly, Fig. \ref{error} shows the total transmit
power and energy efficiency versus the channel uncertainty level $\delta$
when $r=4$ bit/s/Hz. It is observed from Fig. \ref{error}(a) that
the required transmit power of the robust IRS beamforming is higher
than other schemes. This is the price to pay to have a robust design
and to employ passive reflection elements. In any case, it is less
than the ``Non IRS'' case. The energy efficiency (EE) reported in
Fig. \ref{error}(b) is defined as the ratio between the smallest
achievable rate among the users and the total power consumption. The
total power consumption of the IRS schemes is equal to $||\mathbf{F}||_{F}^{2}+NP_{active}+MP_{passive}$
and that of the relay scheme is equal to $||\mathbf{F}||_{F}^{2}+P_{relay}+(N+2M)P_{active}$,
where $P_{relay}$ is the relay transmit power. We set the circuit
power consumption of the active antennas to $P_{active}=10$ mW and
that of the passive antennas as $P_{active}=5$ mW \cite{emil2019}.
Fig. \ref{error}(b) illustrates the high EE performance of the IRS-aided
systems compared with the relay system for the reason of the low circuit
power consumption of the passive elements in the IRS. In addition,
from Fig. \ref{error}(a) and Fig. \ref{error}(b) we come to the
conclusion that only when the reflection efficiency of the reflecting
metasurfaces is high ($\iota$ is nearly 1) and the estimation error
of the reflection channel is small ($\delta$ is less than 0.03),
the IRS can show its advantages of enhancing the spectral and energy
efficiency.

\begin{figure}
\centering \includegraphics[width=2.6in,height=1.8in]{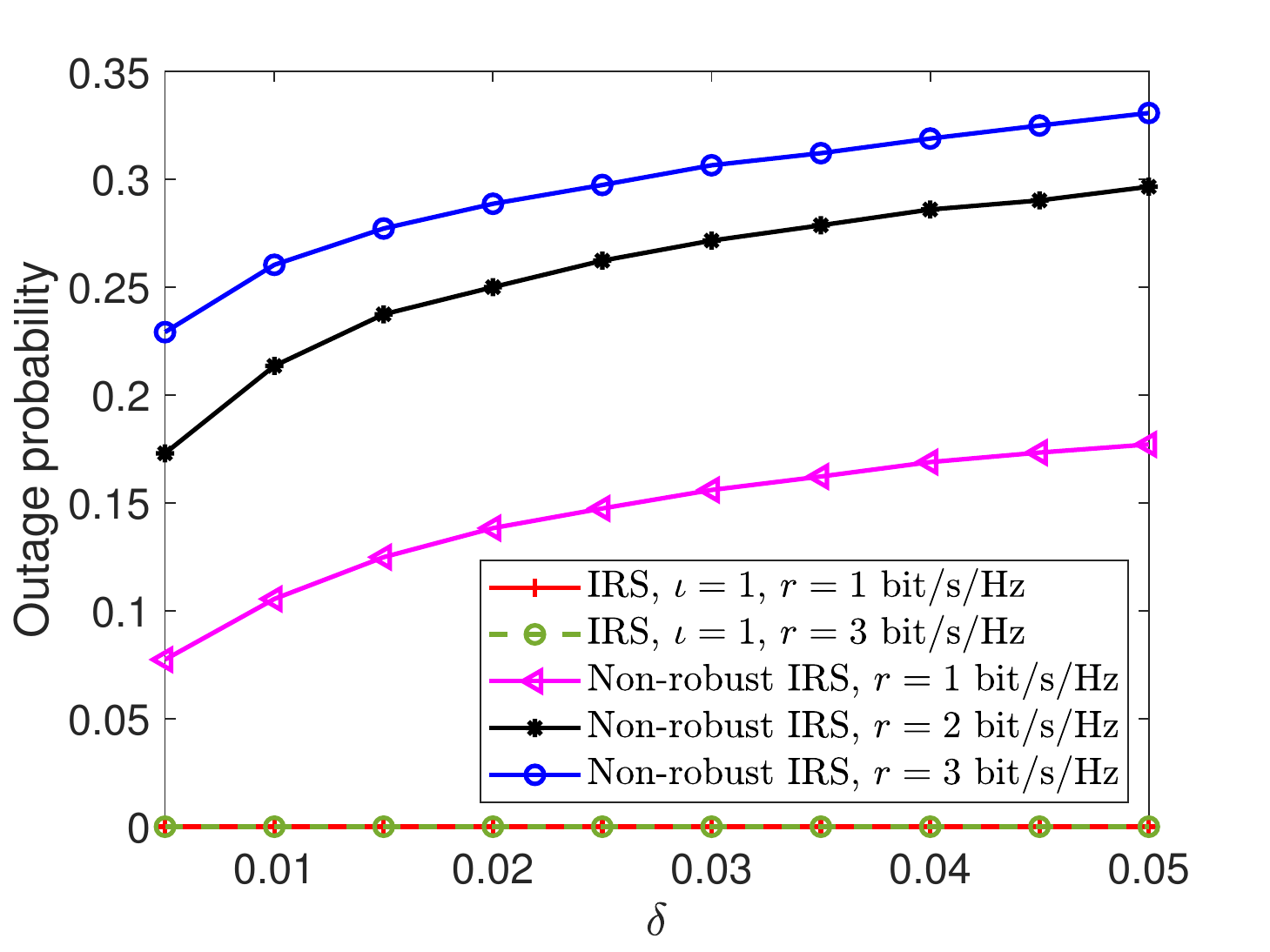}
\caption{Outage probability of rate versus the channel uncertainty
level $\delta$ under $N=6$, $M=16$ and $K=4$.}
\label{rate} 
\end{figure}

Fig. \ref{rate} shows the outage probability of rate
for the nonrobust design. Here, outage probability refers to the probability
that the target rate of at least one user is not satisfied. It is
observed that when the beamforming design ignores the channel error,
the target rate of at least one user is frequently not met, especially
at high value of $r$ or $\delta$. However, our adopted worst-case
robust design method can guatantee no outage happens.

\section{Conclusions}

In this paper, we considered the robust beamforming design for the
IRS-aided MU-MISO system when the CSI is imperfect. The CSI uncertainties
were addressed by using approximation and transformation techniques,
and the non-convex unit-modulus constraints were solved under the
penalty CCP framework. Numerical results demonstrated the robustness
of our proposed algorithm.

 \bibliographystyle{IEEEtran}
\bibliography{bibfile}

\end{document}